\documentclass[proceedings]{JHEP3} % 10pt is ignored!
\conference{Workshop on Integrable Theories, Solitons and Duality}

\usepackage{epsfig,multicol}                    % please use epsfig.

\newbox\mybox
\newcommand\fverb{\setbox\mybox=\hbox\bgroup\verb}
\newcommand\fverbdo{\egroup\medskip\noindent\fbox{\unhbox\mybox}\ }
\newcommand\fverbit{\egroup\item[\fbox{\unhbox\mybox}]}

\newcommand{\be}{\begin{equation}}
\newcommand{\ee}{\end{equation}}
\newcommand{\bea}{\begin{eqnarray}}
\newcommand{\eea}{\end{eqnarray}}
\newcommand{\nn}{\nonumber \\}
\newcommand{\half}{\frac{1}{2}}

\title{NONCOMMUTATIVE SUPERSYMMETRIC THEORIES}

\author{{Victor O. Rivelles} \\ Instituto de F\'{\i}sica \\
 Universidade 
 de S\~{a}o Paulo\\  Caixa Postal 66318, 05315-970, S\~{a}o Paulo, SP,
 Brazil\\ E-mail: rivelles@fma.if.usp.br}

\abstract{We discuss the renormalization properties of noncommutative
  non-gauge supersymmetric field theories.}

\begin{document}

\section{Introduction} 

The main feature of noncommutative field theories is the ultraviolet
(UV) and infrared (IR) 
mixing of divergences \cite{review}. This happens because the
noncommutativity, say of 
$x$ and $y$, $[x,y]=\theta$, at the quantum level entails an
uncertainty $\Delta x \, \Delta y \sim \theta$. Together with the usual
uncertainty relation $\Delta x \, \Delta p_x \sim 1$ we find that
$\Delta y \sim \theta \Delta p_x$, which means that the UV regime in the
$x$-direction produces an IR effect in the $y$-direction and
vice-versa. At the 
field theory level this phenomenon manifests itself as a mixture of UV
and IR divergences already 
at the one loop level \cite{Minwalla}. If we choose to renormalize the
theory in the usual 
way then the remaining IR divergence becomes a source of trouble since
it leads, in general, to non-integrable divergences in higher loop
orders jeopardizing renormalizability. It was then suggested that a
possible way out would be the introduction of supersymmetry
\cite{Ruiz-Ruiz}. Since supersymmetric theories have only 
logarithmic divergences it could be possible that the dangerous UV/IR
mixing could be absent. This was shown at one loop level for the two-point
function of the gauge field \cite{Toumbas}. However, a proof
that this was true in 
general was still lacking. Then the noncommutative Wess-Zumino model
was shown to be free of UV/IR mixing to all loop orders
\cite{Wess-Zumino} providing the
first noncommutative field theory which is fully renormalizable in four 
dimensions. Its low energy properties were studied in detail
\cite{low_energy}. Other 
noncommutative supersymmetric non-gauge theories were also found to be
free of UV/IR mixing. For instance, the supersymmetric nonlinear sigma
model in three dimensions turns out to be renormalizable in the
$1/N$ expansion \cite{Sigma-Model,ssb}. Spontaneous symmetry breaking
also has troubles in the 
presence of noncommutativity \cite{KK,Petriello,Ruiz}. The situation
in three dimensions is improved \cite{ssb} but supersymmetry appears
to play no role in this case. 
Thus, supersymmetry seems to be essential for renormalizability in
noncommutative field theories. However, this turns out to be true only for
non-gauge theories. Supersymmetric gauge theories are still under
intense scrutiny. 

\section{Noncommutative Scalar Field Theory}

Noncommutative field theories are obtained from the commutative ones
just by replacing the ordinary field multiplication by the Moyal product
defined as
\be
\label{4}
\left( \phi_1 \star \phi_2 \right) (x) \equiv \left[ e^{i\frac{1}{2}
    \theta^{\mu\nu} \frac{\partial}{\partial x^\mu}
    \frac{\partial}{\partial y^\nu} } \phi_1(x) \phi_2(y)
\right]_{y=x}.
\ee
Then the noncommutative $\phi^4$ model in $3+1$ dimensions reads as
    \cite{Minwalla}  
\be
S = \int d^4 x \,\, \left( \half \partial_\mu \phi \star \partial^\mu \phi
- \frac{m^2}{2} \phi \star \phi - \frac{g^2}{4!}
\phi\star\phi\star\phi\star\phi \right).
\ee

We now proceed in the standard way. The quadratic terms gives the
propagator. Since the Moyal product has the property 
\be
\int dx \,\, (f \star g)(x) =  \int dx \,\, (g \star f)(x) =
\int dx \,\, f(x) g(x),
\ee
the propagator is the same as in the commutative case. This is a
general property of noncommutative theories: the propagators are not 
modified by the noncommutativity. The vertices, however, are in
general affected by phase factors. In the present case we get, in
momentum space,
 \bea
&& -\frac{g^2}{2} \int d^4x \,\, \phi\star\phi\star\phi\star\phi =
-\frac{g^2}{6} \int dk_1 dk_2 dk_3 dk_4 \,\, \delta 
(k_1+k_2+k_3+k_4) \times \nn
&& [ \cos(\half k_1 \wedge k_2) \cos (\half k_3 \wedge k_4) + 
\cos(\half k_1 \wedge k_3) \cos (\half k_2 \wedge k_4) + \nn
&& \cos(\half k_1 \wedge k_4) \cos (\half k_2 \wedge k_3) ] \,\,
\phi(k_1) \, \phi(k_2) \,  \phi(k_3) \, \phi(k_4). 
\eea

We can now compute the one loop correction for the two-point
function. It is easily found to be 
\be 
\frac{g^2}{3 (2 \pi)^4} \int d^4k \,\, \left( 1 + \half \cos(k \wedge
  p) \right) \frac{1}{k^2 + m^2}.
\ee
The first term is the usual one loop mass correction of the
commutative theory (up to a factor $1/2$) which is quadratically
divergent. The second term is not divergent due to the oscillatory
nature of $\cos(k \wedge 
p)$. This shows that the nonlocality introduced by the Moyal product
is not so bad and leaves us with the same divergence structure of the
commutative theory. This is also a general property of noncommutative
theories \cite{Filk}. To take into account the effect
of the second term we regularize the integral using the Schwinger
parametrization 
\be
\frac{1}{k^2 +m^2} = \int_0^\infty d\alpha \,\, e^{- \alpha (k^2 +
  m^2)} e^{-\frac{1}{\Lambda^2 \alpha}},
\ee
where a cutoff $\Lambda$ was introduced. We find
\be \Gamma^{(2)} = \frac{g^2}{48 \pi^2} [ ( \Lambda^2 - m^2
\ln(\frac{\Lambda^2}{m^2}) + \dots ) + \half ( \Lambda^2_{eff} - m^2
\ln(\frac{\Lambda^2_{eff}}{m^2}) + \dots ) ],
\ee
where
\be
\Lambda^2_{eff} = \frac{1}{\frac{1}{\Lambda^2} + \tilde{p}^2}, \qquad
\tilde{p}^\mu = \theta^{\mu\nu} p_\nu.
\ee
Note that when the cutoff is removed, $\Lambda \rightarrow \infty$,
the noncommutative contribution remains finite providing a natural
regularization. Also $\Lambda^2_{eff} = \frac{1}{\tilde{p}^2}$ which
diverges either when  $\theta \rightarrow 0$ or when $\tilde{p}
\rightarrow 0$. 

The one loop effective action is then
\be
\int d^4p \,\, \half ( p^2 + M^2 + \frac{g^2}{96 \pi^2
(\tilde{p}^2 + 1/\Lambda^2)} - \frac{g^2 M^2}{96 \pi^2}
\ln\left(\frac{1}{M^2(\tilde{p}^2 + 1/\Lambda^2)} \right) + \dots )
  \phi(p) \phi(-p),
\ee
where $M$ is the renormalized mass. Let us take the limits $\Lambda
\rightarrow \infty$ and $\tilde{p} \rightarrow 0$. If we take first
$\tilde{p} \rightarrow 0$ then $\tilde{p}^2 << \frac{1}{\Lambda^2}$
and $\Lambda_{eff}=\Lambda$ showing that we 
recover the effective commutative theory
\be
\int d^4p \,\, \half \left( p^2 + M^{\prime 2} \right) \phi(p)
\phi(-p). 
\ee 
If, however, we take $\Lambda \rightarrow \infty$ then $\tilde{p}^2 >>
\frac{1}{\Lambda^2}$ and $\Lambda^2_{eff} = \frac{1}{\tilde{p}^2}$
and we get  
\be 
\int d^4p \half \left( p^2 + M^2 + \frac{g^2}{96 \pi^2 \tilde{p}^2} -
  \frac{g^2 M^2}{96 \pi^2} \ln\left(\frac{1}{M^2 \tilde{p}^2}\right) +
  \dots \right) \phi(p) \phi(-p),
\ee
which is singular when $\tilde{p} \rightarrow 0$. 
This shows that the limit $\Lambda \rightarrow \infty$ does not
commute with the low momentum limit $\tilde{p} \rightarrow 0$  so that
there is a mixing of UV and IR limits. 

The theory is renormalizable at one loop order if we do not
take $\tilde{p} \rightarrow 0$. What about higher loop orders? Suppose
we have insertions of one loop mass corrections. Eventually we will have
to integrate over small values of $\tilde{p}$ which diverges when
$\Lambda \rightarrow \infty$. Then we find an IR
divergence in a massive theory. This combination of UV and IR
divergences makes the theory non-renormalizable. 

There are also examples of noncommutative theories which are
nonrenormalizable already at
one loop order \cite{Arefeva-complex}. For a complex scalar field with
interaction  $\phi^*\star \phi^* \star \phi \star \phi$ it is found
that the theory is one loop nonrenormalizable while $\phi^*\star \phi
\star \phi^*\star \phi$ gives a  one loop renormalizable  model. 

Then the main question now is the existence of a theory
which is renormalizable to all loop orders. Since the UV/IR mixing
appears at the level of quadratic divergences a candidate theory would
be a supersymmetric one because it does not have such divergences. As
we shall see this indeed happens.  

\section{NONCOMMUTATIVE WESS-ZUMINO MODEL}

The noncommutative Wess-Zumino model in $3+1$ dimensions
\cite{Wess-Zumino} has the action 

\bea
{\cal L}_0 &=&  \frac12 \partial^\mu A \partial_\mu A + \frac12
\partial^\mu B \partial_\mu B
+ \frac12 \overline \psi i\not \! \partial \psi, \\
{\cal L}_m &=& \frac12 F^2+\frac12 G^2
+ m F A + m G B - \half m \overline\psi \psi, \\ 
{\cal L}_g &=&  g (F\star A \star
A- F\star B \star B + G\star A \star B + G \star B \star A - \nn
&& \overline \psi
\star \psi\star A - \overline \psi\star  i\gamma_5 \psi \star B),
\eea
where $A$ and $B$ are bosonic fields, $F$ and $G$ are auxiliary
fields and $\psi$ is a Majorana spinor. The action is invariant under
the usual supersymmetry transformations. The supersymmetry
transformations  are not modified by the 
Moyal product since they are linear in the fields. The elimination of
the auxiliary fields through their equations of motion produces quartic
interactions. In terms of the complex field $\phi=A+iB$ we get $\phi^*
\star \phi^* \star \phi \star \phi$ which is non-renormalizable in the
noncommutative case. This casts doubts about the renormalizability of the
model but as we shall see supersymmetry saves the day. 

As usual, the propagators are not modified by noncommutativity. 
They are given by
\bea
\Delta_{AA}(p) &=& \Delta(p)\equiv\frac{i}{p^2-m^2+i\epsilon},\\
\Delta_{FF}(p) &=& p^2 \Delta(p),\\
\Delta_{AF}(p) &=&\Delta_{FA}(p) = -m \Delta(p), \\
S(p) &=& \frac{i}{\not \! p -m}.
\eea
Taking into account the symmetries, the vertices are 
\bea
F A^2 \quad {\mbox {vextex:}}&& \quad ig \cos(p_1\wedge p_2), \\
F B^2  \quad {\mbox {vextex:}}&& \quad -ig \cos(p_1\wedge p_2),\\
G A B \quad  {\mbox {vertex:}} && \quad 2 ig \cos (p_1\wedge p_2),\\
\overline \psi \psi A\quad {\mbox {vertex:}} &&\quad -ig 
\cos (p_1\wedge p_2),\\ 
\overline \psi \psi B\quad {\mbox {vertex:}} &&\quad -ig\gamma_5 
\cos (p_1\wedge p_2).
\eea
The degree of superficial divergence for a generic 1PI graph
$\gamma$ is then 
\be
d(\gamma)= 4 -  I_{AF} -I_{BF}-N_A-N_B-2 N_F-2N_G - \frac32 N_\psi, 
\ee
where $N_{\cal O}$ denotes the number of external lines
associated to the  field ${\cal O}$ and $ I_{AF}$ and $I_{BF}$
are the numbers of internal lines associated to the mixed
propagators $AF$ and $BF$, respectively. In all cases we will
regularize the divergent Feynman integrals by assuming that a 
supersymmetric regularization scheme does exist.

The one loop analysis can be done in a straightforward way. 
As in the commutative case all tadpoles contributions add up to
zero. We have verified this explicitly. The self-energy of $A$ can
be computed and the divergent part is contained in the integral
\be
16 g^2\int\frac{d^4k}{(2\pi)^4}( 1 + \half \cos(k \wedge p) ) \frac{(p \cdot
  k)^2}{(k^2- m^2)^3}.
\ee
The first term is logarithmically divergent. It differs by a factor 2
from the commutative case. As usual, this 
divergence is eliminated by a wave function renormalization. 
The second term is UV convergent and for small $p$ it behaves as 
$p^2 \ln (p^2/m^2)$ and actually vanishes for $p=0$. Then there is no
IR pole. The same analysis can be carried out for the others
fields. For $F$ we find that the divergent part is
\be
4 g^2 \int \frac{d^4k}{(2\pi)^4 } (1 + \half\cos(k \wedge p) )
\frac{1}{(k^2-m^2)^2}.
\ee
The first term is logarithmically divergent and can also be eliminated
by a wave 
function renormalization. The second term diverges as $\ln(p^2/m^2)$
as $p$ goes to zero. However its multiple insertions is harmless. 
For the fermion field the divergent part is similar to the former
results and needs also a wave
function renormalization. The term containing $\cos(k\wedge p)$ 
behaves as $\not \! p \ln(p^2/m^2)$ and vanishes as $p$ goes to
zero. Therefore, there is no UV/IR mixing in the self-energy as
expected. 

To show that the model is renormalizable we must also look into the
interactions vertices. The $A^3$ vertex has no divergent parts as in
the commutative case. The same happens for the other three point
functions. For the four point vertices no divergence is found as in
the commutative case. Hence, the noncommutative Wess-Zumino model is
renormalizable at one loop with a wave-function renormalization and no
UV/IR mixing.  

To go to higher loop orders we proceed as in the commutative case. We
derived the supersymmetry Ward identities  
for the n-point vertex function. Then we showed that there is a
renormalization prescription which is consistent with the Ward
identities. They are the same as in the commutative case. And finally
we fixed the primitively divergent vertex functions. Then we found
that there is only a common wave function renormalization as in the
commutative case. In general we expect 
\be 
\varphi_R = Z^{-1/2} \varphi, \qquad m_R = Z m + \delta m, \qquad
g_R = Z^{3/2} Z^\prime g.
\ee
At one loop we found { $\delta m = 0$ and $Z^\prime = 1$}. We showed
that this also holds to all orders and no mass renormalization is
needed. 

Being the only consistent noncommutative quantum field theory in $3+1$
dimensions known so far it is natural to study it in more detail. As a
first step in this direction we considered the non-relativistic limit
of the noncommutative Wess-Zumino model \cite{low_energy}. We found
the low energy 
effective potential mediating the fermion-fermion and boson-boson
elastic scattering in the non-relativistic regime. Since
noncommutativity breaks Lorentz invariance we formulated the theory in
the center of mass frame of reference where the dynamics simplifies
considerably. For the fermions we found that the potential is
significantly changed by the noncommutativity while no modification
was found for the bosonic sector. The modifications found give rise to
an anisotropic differential cross section. 

Subsequently the model was formulated in superspace and again found to
be renormalizable to all loop orders \cite{Bichl}. The one and two
loops contributions to the effective action in superspace were also
found \cite{Buchbinder}. The one loop Kahlerian effective potential
does not get modified by noncommutativity and the two loops nonplanar
contributions to the Kahlerian effective potential are leading in the
case of small noncommutativity \cite{Buchbinder}.

\section{Noncommutative Gross-Neveu and  Nonlinear Sigma Models}

Another model where non-renormalizability is spoiled by the
noncommutativity is the $O(N)$ Gross-Neveu model. The commutative
model is 
perturbatively renormalizable in $1+1$ dimensions and $1/N$
renormalizable in $1+1$ and $2+1$ dimensions. In both cases it
presents dynamical mass generation. It is described by the Lagrangian 
\begin{equation}
{\cal L}=\frac{i}2\overline \psi_i \not \!\partial \psi_i +\frac
{g}{4N}(\overline \psi_i \psi_i)(\overline \psi_j \psi_j),\label{1}
\end{equation}
where $\psi_i, i=1,\ldots N$, are two-component Majorana
spinors. Since it is renormalizable in the $1/N$ expansion in $1+1$
and $2+1$ dimensions we will consider both cases. As usual, we
introduce an auxiliary field $\sigma$ and the Lagrangian turns into 
\begin{equation}
{\cal L} =\frac{i}2\overline \psi_i \not \!\partial \psi_i -\frac{\sigma}2 
(\overline \psi_i \psi_i)- \frac{N}{4g}\sigma^2.\label{2}
\end{equation}
Replacing $\sigma$ by $\sigma+M$ where $M$ is the VEV of the
original $\sigma$ we get the gap equation (in Euclidean space) 
\begin{equation}
 \frac{M}{2g}-\int \frac{d^Dk}{(2\pi)^D}\frac{M}{k^{2}_{E}+M^2}=0.\label{5}
\end{equation}
To eliminate the UV divergence we need to renormalize the coupling
constant by 
\begin{equation}
\frac{1}{g}= \frac{1}{g_R} + 2 \int \frac{d^Dk}{(2\pi)^D}\frac{1}{k^{2}_{E}
+\mu^2}. \label{6}
\end{equation}
In $2+1$ dimensions we find 
\begin{equation}
\frac{1}{g_R}= \frac{\mu-|M|}{2\pi}, \label{7}
\end{equation}
and therefore only for $-\frac{1}{g_R}+\frac{\mu}{2\pi}>0$ it is possible to
have $M\not =0$, otherwise $M$ is necessarily zero. 
No such a restriction exists in $1+1$ dimensions. In any case, 
we will focus only in the massive phase.
The propagator for $\sigma$ is proportional to the inverse of the
following expression
\begin{equation}
 -\frac {iN}{2g}-  iN \int \frac{d^Dk}{(2\pi)^D} \frac{k\cdot (k+p) + 
M^2}{(k^2-M^2)[(k+p)^2-M^2]}, \label{8}
\end{equation}
which is divergent. Taking into account the gap equation the above
expression reduces to 
\begin{equation}
\frac{(p^2-4M^2)N}2\int\frac{d^Dk}{(2\pi)^D} \frac{1}{(k^2-M^2)[(k+p)^2-M^2]},
\label{10}
\end{equation}
which is finite. Then there is a fine tuning which is responsible for
the elimination of the divergence and which might be absent in the
noncommutative case due to the UV/IR mixing. 

The noncommutative model is defined by \cite{Sigma-Model} 
\begin{equation}
S_{GN}=\int d^Dx \left [\frac{i}2\overline \psi \not \!\partial \psi -
  \frac{M}2\overline \psi 
\psi-\frac12\sigma \star(\overline \psi\star
\psi)- \frac{N}{4g}\sigma^2- \frac{N}{2g}M\sigma\right ].\label{11}
\end{equation} 
Elimination of the auxiliary field results in a four-fermion
interaction of the type $\overline \psi_i\star\psi_i\star\overline
\psi_j\star\psi_j$. However a more general four-fermion interaction
may involve a term like $\overline \psi_i\star\overline
\psi_j\star\psi_i\star\psi_j$. This last combination does not have a
simple $1/N$ expansion and we will not consider it. The Moyal product
does not affect the propagators and the trilinear vertex acquires a
correction of $\cos(p_1\wedge p_2)$ with regard to the commutative
case. Hence the gap equation is not modified, while the propagator for
the $\sigma$ is now proportional to the inverse of
\begin{equation}
 -\frac {iN}{2g}- N \int \frac{d^Dk}{(2\pi)^D} \cos^2(k\wedge p)
\frac{k\cdot (k+p) + 
M^2}{(k^2-M^2)[(k+p)^2-M^2]}. \label{12}
\end{equation}
Now the divergent part is no longer canceled and this turns the model
into a nonrenormalizable one. 

On the other side, the nonlinear sigma model also presents troubles in
its noncommutative version. The noncommutative model is described by 
\begin{equation}\label{121}
{\cal L} =-\frac12\varphi_i (\partial^2 + M^2)\varphi_i + \frac12
\lambda\star \varphi_i \star \varphi_i - \frac{N}{2g}\lambda,
\end{equation}
where $\varphi_i$, $i=1,\ldots , N$, are real scalar fields, $\lambda$
is the auxiliary field and $M$ is the generated mass. The leading
correction to the $\varphi$ self-energy is
\begin{equation}\label{122}
-i\int \frac{d^2 k}{(2\pi)^2} \frac{\cos^2 (k\wedge
 p)}{(k+p)^2-M^2}\Delta_\lambda(k), 
\end{equation}
where $\Delta_\lambda$ is the propagator for $\lambda$. As for the
case of the scalar field this can be decomposed as a sum of a
quadratically divergent part and a UV finite part. Again there is the
UV/IR mixing destroying the $1/N$ expansion. 

\section{Noncommutative Supersymmetric Nonlinear Sigma Model}

The Lagrangian for the commutative supersymmetric sigma model is given
by 
\begin{equation}
{\cal L} =\frac12 \partial^\mu \varphi_i \partial_\mu \varphi_i +
\frac{i}{2} \overline \psi_i \not \! \partial \psi_i + \frac12 F_i F_i
+ \sigma \varphi_i F_i + \frac12\lambda 
\varphi_i \varphi_i  - \frac12\sigma \overline \psi_i \psi_i -
\overline \xi \psi_i \varphi_i - \frac{N}{2g}\lambda,\label{13} 
\end{equation}
where $F_i$, $i=1,\ldots, N$, are auxiliary fields. Furthermore,
$\sigma,\lambda$ and $\xi$ are the Lagrange multipliers which
implement 
the supersymmetric constraints. After the change of variables 
$\lambda\rightarrow \lambda + 2 M \sigma$, $F\rightarrow F-M\varphi$
where $M=<\sigma>$, and the shifts $\sigma\rightarrow \sigma +M$ and
$\lambda\rightarrow \lambda + \lambda_0$, where $\lambda_0=<\lambda>$,
we arrive at a more symmetric form for the Lagrangian 
\begin{eqnarray}
{\cal L} &=& - \frac12 \varphi_i ( \partial^2+M^2) \varphi_i + \frac{1}{2} 
\overline \psi_i (i\not \! \partial - M)\psi_i + \frac12 F_i^2+
 M^2 \varphi_i^2+ \frac12\lambda_0 \varphi_i^2   \nonumber\\ 
&\phantom a& + \frac12\lambda \varphi_i^2 +\sigma  \varphi_i F_i
 - \frac12\sigma \overline \psi_i \psi_i  - \overline \xi \psi_i
 \varphi_i- \frac{N}{2g}\lambda -\frac{N}{g}M\sigma.\label{15} 
\end{eqnarray}
Now supersymmetry requires $\lambda_0=-2M^2$ and the gap equation is 
\begin{equation}
\int \frac{d^D k}{(2\pi)^D} \frac{i}{k^2-M^2}= \frac{1}{g},\label{16}
\end{equation}
so a coupling constant renormalization is required. We now must
examine whether the propagator for $\sigma$ depends on the this
renormalization.  We find that the two point function for $\sigma$ is
proportional to the inverse of
\begin{equation}
 \frac{(p^2-4M^2)N}2\int\frac{d^Dk}{(2\pi)^D}
 \frac{1}{(k^2-M^2)[(k+p)^2-M^2]}\,\,\label{18},
\end{equation}
which is identical to the Gross-Neveu case. Notice that the gap equation
was not used. The finiteness of the above expression is a consequence
of supersymmetry. 

The noncommutative version of the supersymmetric nonlinear sigma model
is given by \cite{Sigma-Model}
\begin{eqnarray}
{\cal L} &=&  
- \frac12\varphi_i (\partial^2+M^2) \varphi_i + \frac{1}{2} \overline
\psi_i (i\not \! \partial -M)\psi_i + \frac12 F_i^2 +
\frac{\lambda}{2}\star\varphi_i \star\varphi_i  \nonumber \\
& \phantom a & - \frac12 F_i \star (\sigma\star\varphi_i +\varphi_i
\star\sigma)-\frac12 \sigma\star\overline\psi_i \star \psi_i  
-\frac12 (\bar\xi\star\psi_i \star\varphi_i
+\bar\xi\star\varphi_i\star\psi_i ) \nonumber\\
&\phantom a &-\frac{N}{2g}\lambda - \frac{N M\sigma}{g}.
\label{19}
\end{eqnarray}
Notice that supersymmetry dictates the form of the trilinear
vertices. Also, the supersymmetry transformations are not modified by
noncommutativity since they are linear and no Moyal products are
required. 

The propagators are the same as in the commutative case. The vertices
have cosine factors due to the Moyal product 
%\begin{mathletters}
\begin{eqnarray}\label{20}
\lambda \varphi^2 \quad {\mbox {vertex:}}&& \quad \frac{i}2 
\cos(p_1\wedge p_2), \\
\sigma \varphi F \quad  {\mbox {vertex:}} && \quad - i \cos (p_1\wedge p_2),\\
\overline \psi \psi \sigma \quad {\mbox {vertex:}} &&\quad -\frac{i}2 
\cos (p_1\wedge p_2),\\ 
\overline \xi \psi \varphi \quad {\mbox {vertex:}} &&\quad - i 
\cos (p_1\wedge p_2).
\end{eqnarray}
%\end{mathletters}
We again consider the propagators for the Lagrange multiplier
fields. Now the $\sigma$ propagator is modified
by the cosine factors and is proportional to the inverse of 
\begin{equation}
\frac{(p^2-4M^2)N}2\int\frac{d^Dk}{(2\pi)^D}
 \frac{\cos^2(k\wedge p)}{(k^2-M^2)[(k+p)^2-M^2]}\label{21}.
\end{equation}
It is well behaved both in UV and IR regions. The propagators for
$\lambda$ and $\xi$ are proportional to the inverse of
\begin{equation} 
\frac{N}{2}\int \frac{d^D k}{(2\pi)^D}
\cos^2(k\wedge p)\frac{1}{[(k+p)^2-M^2] [k^2-M^2]},
\label{22}
\end{equation}
and
\begin{equation}
{N} \frac{(\not \! p + 2 M)}2\int \frac{d^D k}{(2\pi)^D}
{\cos^2(k\wedge p)} \frac{1}{[(k+p)^2-M^2][k^2-M^2]},\label{23}
\end{equation}
respectively. They are also well behaved in UV and IR regions. 

The degree of superficial divergence for a generic 1PI graph $\gamma$ is
\begin{equation}
d(\gamma)= D - \frac{(D-1)}2N_\psi- \frac{(D-2)}2 N_\varphi-\frac{D}2
N_F- N_\sigma- \frac32 N_\xi- 2 N_\lambda,\label{24}
\end{equation}
where $N_{\cal O}$ is the number of external lines associated to the 
field ${\cal O}$. Potentially dangerous diagrams are those contributing
to the self--energies of the $\varphi$ and $\psi$ fields since, in principle,
they are quadratic and linearly divergent, respectively.
For the self-energies of $\varphi$ and $\psi$  we find that they
diverge logarithmically and they can be removed by a wave function
renormalization of the respective field. The same happens for the
auxiliary field $F$. The renormalization factors for them are the same
so supersymmetry is preserved in the noncommutative theory. 
This analysis can be extended to the n-point functions. In $2+1$
dimensions we find nothing new showing the renormalizability of the
model at leading order of $1/N$. However, in $1+1$ dimensions there
some peculiarities. Since the scalar field is dimensionless in $1+1$
dimensions any graph involving an arbitrary number of external
$\varphi$ lines is quadratically divergent. In the four-point function
there is a partial cancellation of divergences but a logarithmic
divergence still survives. The counterterm needed to remove it can not
be written in terms of $\int d^2 x \,\,\varphi_i \star \varphi_i \star
\varphi_j\star\varphi_j$ and $\int d^2 x \,\,
\varphi_i\star\varphi_j\star\varphi_i\star\varphi_j$. A possible way 
to remove this divergence is by generalizing the definition of 1PI
diagram. However the cosine factors do not
allow us to use this mechanism which casts doubt about the
renormalizability of the noncommutative supersymmetric $O(N)$
nonlinear sigma model in $1+1$ dimensions. 

The noncommutative supersymmetric nonlinear sigma model can also be
formulated in superspace \cite{sigma-superspace}. There it is possible
to go beyond the sub-leading order in $1/N$. It is then possible to
show that model is renormalizable to all orders of 1/N and explicitly
verify that it is asymptotically free \cite{sigma-superspace}.   

\section{Spontaneous Symmetry Breaking in Noncommutative Field Theory}

Having seen the important role supersymmetry plays in noncommutative
models it is natural to go further. Spontaneous symmetry breaking and
the Goldstone theorem are essential in the 
standard model and the effect of noncommutativity in this setting
deserves to be fully 
understood. In four dimensions it is known that spontaneous symmetry
breaking can occur for the $U(N)$ model but not for the $O(N)$ unless
$N = 2$.  The $O(2)$ case was analyzed in detail
\cite{Petriello,Ruiz} and 
the results for the $U(N)$ case have been extended to two loops
\cite{Liao}. Going to higher loops requires an IR regulator which can
no longer be removed\cite{Sarkar}. Due to these troubles we will
consider three dimensional models. 

Let us consider the three-dimensional action \cite{ssb} 
\begin{eqnarray}
\label{action1}
S&=&\int d^3 x \Big[-\frac{1}{2}\phi_a\Box\phi_a+\frac{\mu^2}{2}\phi_a\phi_a
\nonumber\\&-&
\frac{g}4 \Big ( l_1 \phi_a*\phi_a*\phi_b*\phi_b
+l_2\phi_a*\phi_b*\phi_a*\phi_b\Big)
\nonumber\\
&-&
\frac{\lambda}{6}\Big(h_1\phi_a*\phi_a*\phi_b*\phi_b*\phi_c*\phi_c+
h_2\phi_a*\phi_a*\phi_b*\phi_c*\phi_c*\phi_b+\nonumber\\&+&
h_3\phi_a*\phi_a*\phi_b*\phi_c*\phi_b*\phi_c+h_4\phi_a*\phi_b*\phi_c*\phi_a*\phi_b*\phi_c+
\nonumber\\&+&
h_5\phi_a*\phi_b*\phi_c*\phi_a*\phi_c*\phi_b
\Big)
\Big],
\end{eqnarray}
where $l_1,l_2,h_1,h_2\ldots h_5$ are real numbers satisfying the conditions
$l_1+l_2=1$ and $h_1+h_2+\ldots+h_5=1$, so that there are two quartic
and five sextuple independent interaction couplings. The potential has
a minimum for $\phi_a\phi_a=a^2$ with 
\begin{equation}\label{one}
a^2= \frac{1}{2\lambda} \left ( -g + \sqrt{g^2+ 4\mu^2\lambda}\right ).
\end{equation}
As usual we introduce the field $\pi_i$ and $\sigma$ having vanishing
expectation value. The action then becomes 
\begin{eqnarray}
\label{action4}
S&=&\int d^3 x
\Big\{-\frac{1}{2}\pi_i\Box \pi_i
-\frac{1}{2}\sigma(\Box-m^2)\sigma-(2\lambda a^3+ g a)\sigma*\pi_i*\pi_i-(\frac{10}{3} \lambda a^3+g a) \sigma*\sigma*\sigma
\nonumber\\&-&
\Big[(\frac{\lambda}{6}\alpha a^2+\frac{g}4l_1)\pi_i*\pi_i*\pi_j*\pi_j+
(\frac{\lambda}{6}(3-\alpha) a^2+\frac{g}4l_2)
\pi_i*\pi_j*\pi_i*\pi_j\nonumber \\
&+&
(\frac{\lambda}{6} a^2\beta +\frac{g}2 l_1) \sigma*\sigma*\pi_i*\pi_i+(\frac{\lambda}{6} a^2(18
-\beta)+ \frac{g}2 l_2)\sigma*\pi_i*\sigma*\pi_i\nonumber\\&+&
+(\frac{5}{2} \lambda a^2
+\frac{g}4) \sigma*\sigma*\sigma*\sigma
\Big]+\ldots\Big \},\label{ac0}
\end{eqnarray}
where $m^2= 4\mu^2-2 g a^2= 4\lambda a^4+2g a^2$, 
the dots denote terms of fifth and sixth order in the fields,
$\alpha=3h_1+2(h_2+h_3)+h_5$, 
and $\beta=18h_1+14h_2+12h_3+6h_4+8h_5$. 
Notice that condition (\ref{one}) implies that the pions 
are massless in the tree approximation, in agreement with the
Goldstone theorem. 

We then find the propagators 
\begin{eqnarray}
&&<\sigma(p_1)\sigma(p_2)>= (2\pi)^d \delta(p_1+p_2) \frac{i}{p_{1}^2-m^2},\\
&&<\pi_i(p_1)\pi_j(p_2)>= (2\pi)^d \delta(p_1+p_2)
\frac{i\delta_{ij}}{p_{1}^{2}},
\end{eqnarray}
and the vertices present the usual phase factors 
\begin{eqnarray}
& &\pi_i(p_1)\pi_j(p_2)\pi_k(p_3)\pi_l(p_4) 
\to -i\rho_1 [\cos(p_1\wedge p_2)\cos(p_3\wedge p_4)\delta_{ij}\delta_{kl}\nonumber\\
&&+
\cos(p_1\wedge p_3)\cos(p_2\wedge p_4)\delta_{ik}\delta_{jl}+
\cos(p_1\wedge p_4)\cos(p_2\wedge p_3)\delta_{il}\delta_{kj}]
\nonumber\\&-&
i\rho_2[\cos(p_1\wedge p_3+p_2\wedge
p_4)\delta_{ij}\delta_{kl}
+\cos(p_1\wedge p_2+p_3\wedge
p_4)]\delta_{ik}\delta_{jl}\nonumber\\& &+
\cos(p_1\wedge p_2+p_4\wedge p_3)\delta_{il}\delta_{kj}],\\
[0.3cm]
& &\pi_j(p_1)\pi_j(p_2)\sigma(p_3)\sigma(p_4) \to 
-i[\rho_3\cos(p_1\wedge p_2)\cos(p_3\wedge p_4)+\nonumber\\&+&
\rho_4\cos(p_1\wedge p_3+p_2\wedge p_4)],\\
[0.3cm]
& &\sigma(p_1)*\pi_i(p_2)*\pi_i(p_3) \to
-i(4\lambda a^3+2 g a)\cos(p_2\wedge p_3),
\end{eqnarray}
where $\rho_1=\frac{4\lambda}{3}a^2 \alpha+2 gl_1$,
$\rho_2=(3-\alpha)\frac{4\lambda}{3}a^2+ 2 g l_2$,
$\rho_3=\frac{2\lambda}{3}a^2\beta+ 2 g l_1$ and
$\rho_4=\frac{2\lambda}{3}a^2(18-\beta)+2gl_2$.

The gap equation receives no contribution from noncommutativity while
the one loop corrections to the pion mass have divergences both, in the
planar and non-planar sectors. Eliminating the UV divergence in the
planar sector also eliminates the UV/IR mixing in the non-planar
sector. It is also fortunate that it leads to an analytic behavior in
the IR so that the mass corrections vanish for $p=0$. This mechanism
does not appears in the four dimensional case. 

The two point function for $\sigma$ is also analytic in the IR leading
to a relation among the parameters. The divergences in the higher
point functions can also be eliminated. Therefore, we have shown that
this $O(N)$ model is renormalizable at one loop for any $N$
\cite{ssb}, in contradistinction to the four dimensional case where
$N$ must be equal to 2.  

A supersymmetric version of this model can be formulated in
superspace. Again, the gap equation is not affected by
noncommutativity. The mass corrections for the pion two point function
are UV finite and
free of UV/IR mixing as expected. It also vanishes for
$p=0$. Supersymmetry does not appear to be important in
this situation.

\section{Conclusions}

We have shown that it is possible to build consistent quantum field
theories in noncommutative space. Supersymmetry is
an essential ingredient for renormalizability. The models studied here
do not involve gauge fields and this considerably simplifies the
situation. All vertices are deformed in the same way by the Moyal
product and this was essential to analyze the amplitudes. With gauge
fields the situation is much more complicated because the vertices are
deformed in different ways. However,
supersymmetric gauge theories may still have a better behavior. The
analysis of spontaneous symmetry breaking in three dimensions revealed 
that it is possible to implement the Goldstone theorem in
noncommutative theories. Supersymmetry seems to play no essential 
role in this case.

\acknowledgments
This work was done in collaboration with H. O. Girotti, M. Gomes, 
A. J. da Silva and A. Petrov. It was partially supported by Conselho
Nacional de Desenvolvimento Cient\'\i fico e Tecnol\'ogico (CNPq), and
PRONEX under contract CNPq 66.2002/1998-99.

\end{document}